\documentclass[sigplan,screen,anonymous=false]{acmart}

\usepackage[utf8]{inputenc}
\usepackage{graphicx} 
\usepackage{booktabs}
\usepackage{duckuments}
\usepackage{listings}
\usepackage{xcolor}
\usepackage{mdframed}
\usepackage{soul}

\setcopyright{none}
\copyrightyear{2024}
\acmYear{2024}
\acmDOI{}
\renewcommand\footnotetextcopyrightpermission[1]{}
\acmConference{}{}{}
\newcommand{\tool}[0]{\texttt{dafny-annotator}}

\title[\tool: AI-Assisted Verification of Dafny Programs]{\tool:\\ AI-Assisted Verification of Dafny Programs}
\author[3]{Gabriel Poesia$^1$ \ Chloe Loughridge$^2$ \ Nada Amin$^2$}

\affiliation{%
  \institution{$^1$Stanford University \hspace{1cm} $^2$Harvard University}
  \country{ }
}
\email{poesia@cs.stanford.edu}
\email{cloughridge@college.harvard.edu}
\email{namin@seas.harvard.edu}
\date{October 2024}

\lstnewenvironment{Dafny}[1][]
      {\lstset{language=Java}\lstset{escapeinside={(*@}{@*)},
       morekeywords={ensures, invariant, nat, method, returns, assert, decreases},
  keywordstyle=\color{blue},
  commentstyle=\color{green!60!black},
  numbers=left,
  numberstyle=\tiny,
  breaklines=true,
  escapeinside={(*@}{@*)},
           showstringspaces=false,
           keywordstyle=\itshape\color{blue},
           stringstyle=\color{maroon},
        commentstyle=\color{black},
        rulecolor=\color{black},
        xleftmargin=0pt,
        xrightmargin=0pt,
        aboveskip=\medskipamount,
        belowskip=\medskipamount,
               backgroundcolor=\color{white}, #1
    }}{}

\begin{document}
\begin{abstract}
    Formal verification has the potential to drastically reduce software bugs, but its high additional cost has hindered large-scale adoption. While Dafny presents a promise to significantly reduce the effort to write verified programs, users are often required to provide logical annotations to aid the verifier. Here, we explore using a combination of Large Language Models and search to build \texttt{dafny-annotator}: a tool that adds logical annotations to a Dafny method until the verifier can prove it correct. On a test set from the DafnyBench collection of programs, greedy search guided by LLaMa 3.1 8B successfully annotates only 15.7\% of the methods. Since this data-driven approach is hindered by the lack of large-scale training data, we propose a method for open-ended synthesis of new Dafny programs in a flexible pipeline where LLMs formulate high-level ideas, implement them, and incrementally propose changes to existing programs, which Dafny validates. This gives us a synthetic dataset, DafnySynth, which we use to augment DafnyBench for training. Fine-tuning on both datasets boosts LLaMa 8B's success rate to 50.6\% --- significantly better than the base model, or training on either dataset alone. Our results suggest a path towards capable AI assistants for languages that don't yet have large-scale human-generated examples. In turn, such assistants might reduce friction for users and ultimately drive adoption.
\end{abstract}

\maketitle
\section{Introduction}

Software bugs remain a persistent and costly problem in the technology industry, with estimates that poor software quality cost the US economy \$2.08 trillion in 2020 alone \cite{krasner2021cost}. Formal verification has the potential to drastically improve software quality: in verified software, the error surface (modulo performance considerations) is reduced to its \emph{specification}, while the \emph{implementation} is proved to adhere to its specified behavior. However, practical adoption of formal software verification has still been limited to specialized projects where errors are exceptionally costly, due to the non-trivial cost added by verification itself.

This extra cost stems from both (a) formalizing the \emph{specification} of the program and (b) creating \emph{verification} artifacts (e.g., formal proofs) that a formal verifier needs to validate the program's implementation. In many widely used theorem provers, like Coq \cite{coq} and Isabelle \cite{isabelle}, the cost of verifying the program is particularly high, as users must write correctness proofs in addition to the program itself. Though users can partially automate this process using \emph{tactics} --- reusable proof producing procedures ---, the proof scripts needed to verify realistic programs are still often larger than the program itself. In contrast, Dafny's design \cite{dafny} proposes to largely automate verification, relying on SMT (Satisfiability Modulo Theory) encodings and solvers. Instead of complete proofs, users only need to provide logical annotations (e.g., assertions, or loop invariants) to guide the verifier when it doesn't succeed, and sometimes prove helper lemmas. While this already substantially reduces user input, adding these annotations is still non-trivial, requiring knowledge of formal semantics and logic than most software developers have not been exposed to, as well as bespoke intuition on what triggers the automated verification.

In this work, we introduce \tool\footnote{Our code and data are available at \url{https://github.com/metareflection/dafny-annotator}}, a tool that leverages Large Language Models (LLMs) and search to automatically add logical annotations to Dafny methods. Concretely, given a Dafny method, with its formal specification and implementation, \tool{} will use an LLM to make proposals of annotations to add to the program, automatically localize where to insert the annotations, and iterate this procedure until Dafny can prove the method's specification (or we reach a maximum number of trials).

We use DafnyBench \cite{loughridge2024dafnybench}, a recent benchmark of 1326 standalone Dafny programs collected from Github, to evaluate \tool. Specifically, we remove the human-written annotations from a held-out test set of methods, and test whether \tool can annotate each resulting method until Dafny verifies it. We use LLMs from the LLaMa family to make annotation proposals, and evaluate their success rate at eventually arriving at fully verified methods. The base LLaMa 8B model only obtains a success rate of 15.7\% at this task. We then describe a simple method for fine-tuning a base LLM using existing annotated programs for this task of guiding an annotator: fine-tuning on a subset of DafnyBench boosts success rate to 20.5\%.

While data-driven methods improve given more data, the number of Dafny programs available for training is extremely small compared to popular languages like Python, which are featured in tens of millions of open-source repositories. To improve \tool{}, we thus propose a method for open-ended synthesis of new programs, capable of, in principle, generating, and progressively adding complexity to, an unbounded number of programs. We use our method to generate DafnySynth, a synthetic dataset of size comparable to DafnyBench. Combining both training sets improves \tool{}'s success rate to 50.7\%, suggesting we're able to generate high-quality training data with this synthetic procedure. We discuss implications of these results and lay out a path for improving AI assistants for languages like Dafny, where (a) existing human data might be insufficient for training assistants, and (b) capable AI assistants might have a positive impact on its adoption.

\section{Methods}
\label{sec:method}

\tool{} aims to automate the addition of logical annotations to Dafny programs using a combination of Large Language Models (LLMs) and search strategies. We divide our method into three main components: an LLM-guided search to annotate methods, a fine-tuning procedure using an existing training set of annotated programs, and a pipeline to synthesize a potentially unbounded number of programs for training.

\subsection{LLM-Guided Greedy Search}

To automatically annotate Dafny methods, we employ a simple greedy search for annotations that the Dafny verifier accepts. Given a Dafny program with its implementation and specification but lacking necessary annotations, we proceed as follows:

\begin{enumerate} \item \textbf{Annotation Proposal:} We prompt the LLM to generate $K$ candidate annotations. The prompt includes the current program, up to the method we are annotating, with its existing annotations. We only require the LLM to output the annotation, but not the location at which it should be inserted.

\item \textbf{Insertion:} For each candidate annotation, we attempt to insert it at all syntactically valid locations within the method (e.g., for loop invariants, we try adding it to each of the loops; for assertions, we add it after every statement). We try these options in parallel for efficiency. This makes the LLM's task easier, since it has less to predict.

\item \textbf{Greedy Selection:} We follow a greedy strategy by keeping the first annotation and program location that Dafny accepts. If none of the candidates verify in any location in the current program, we keep the existing program at the end of this step.
\end{enumerate}

\tool{} repeats these steps up to a maximum number of iterations, stopping early if Dafny can prove the method's post-condition after adding an annotation.

\subsection{Fine-Tuning the LLM for Annotation Generation}

While general-purpose LLMs like LLaMA~\cite{llama} can generate plausible annotations, their performance improves significantly when fine-tuned on task-specific data. To this end, we create a fine-tuning dataset for this task of proposing annotations for a given program by leveraging existing annotated Dafny programs.

Concretely, in the search algorithm above, we sample annotations from the LLM given the current program. Thus, our fine-tuning dataset should consist of pairs of $(p_i, a_i)$ programs and an annotation to be added to that program. Given an annotated program $p$, we thus construct these training pairs by removing the last annotation $a$ (e.g., assertion, invariant, or decreases clause) from the program $p$, yielding a program $p_{-a}$ missing one annotation. The pair $(p_{-a}, a)$ thus becomes a training example. We repeat this procedure with program $p_{-a}$ until there are no more annotations. Each of these pairs becomes one training example, as shown in Figure~\ref{fig:example-prediction}.

    \lstset{
      language=Java,
      basicstyle=\ttfamily,
      keywordstyle=\color{blue},
      commentstyle=\color{green!60!black},
      numbers=left,
      numberstyle=\tiny,
      frame=single,
      morekeywords={ensures, invariant, nat, method, returns, assert, decreases},
      escapeinside={(*@}{@*)}
    }

\begin{figure}
\centering
\footnotesize
\begin{Dafny}
(*@\normalfont\textit{
Given each Dafny program, propose an assertion, invariant or decreases statement in order to verify the program.
\ \\
Program:
}@*)
method maxArray(a: array<int>) returns (m: int)
  requires a.Length >= 1
  ensures forall k :: 0 <= k < a.Length ==>
    m >= a[k]
  ensures exists k :: 0 <= k < a.Length &&
    m == a[k]
{
  m := a[0];
  var index := 1;
  while (index < a.Length)
     decreases a.Length - index
  {
    m := if m>a[index] then  m else a[index];
    index := index + 1;
  }
}
(*@\normalfont\textit{Annotation:}@*) (*@\hl{invariant 0 <= index <= a.Length}@*)
\end{Dafny}
    \caption{Example of an annotation prediction prompt using an LLM. This format is used by \tool{} to sample candidate annotations from an LLM, which it then tries to insert in the program. For training \tool{}, we extract examples in this format from existing annotated programs, where we remove one annotation and train the LLM to predict it (\hl{highlighted} portion of the example).}
    \label{fig:example-prediction}
\end{figure}

We collect a training set for this annotation prediction problem from DafnyBench~\cite{loughridge2024dafnybench}, the largest existing dataset of Dafny programs. DafnyBench consists of 1326 stand-alone Dafny programs collected from Github, deduplicated and filtered for validity (i.e., Dafny verifies all of them). For each program, we systematically remove all annotations, and simulate their reinsertion into the program one by one, as described above. These examples are formatted as strings as shown in Figure~\ref{fig:example-prediction}, on which we can fine-tune the LLM using the standard next-token prediction objective. We use Supervised Fine-Tuning (SFT), meaning that we mask out the tokens corresponding to the prompt (e.g., the existing program) in the loss. This focuses training on the generation of the annotations conditioned on the program, rather than also training the model to predict the program tokens.

\subsection{Edit Graph for Synthetic Dataset Generation}

As we show in Section~\ref{sec:experiments}, even modest fine-tuning can drastically improve the performance of \tool. However, the scarcity of large-scale datasets of Dafny programs poses a significant limitation for training data-intensive models like LLMs. We note that this presents a challenge to the Programming Languages community to propose new tools more broadly: as experienced and new developers alike become used to AI assistance, new languages, even if they might contain significant design advantages over existing ones, can become less appealing due to existing AIs being unfamiliar with them. To overcome this challenge, we need to be able to train AIs to assist human users even for new languages, without relying on decades of accumulated human-written programs for training (as is the case for languages like Python and Java).

To address this, we develop a general methodology for synthesizing an arbitrarily large number of valid programs, bootstrapping from LLMs imperfect knowledge of Dafny and their ability to do in-context learning. We first instantiate this method in Dafny, relying on the Dafny verifier to ensure correctness.

Our method is based on the simple observation that large programs are written as a series of much smaller edits to existing programs. An initial code repository might start with a simple prototype of some core functionality, and evolve from there, diff by diff --- over time, these changes compound towards a large project. We will use LLMs to likewise start by writing a simple program, then iteratively propose small diffs to this program. At each step, we can rely on Dafny to ensure that the changes keep programs syntactically and semantically valid.

Concretely, we propose to create an \emph{Edit Graph}, where nodes have a type and some content, which can be turned into new nodes by a series of \emph{Editors}. Edges represent the relationship that a node was derived from another. When an Editor is scheduled to run, it (a) selects nodes from the existing graph that it is able to edit, and (b) proposes new nodes based on the existing ones. We initialize the graph with a single node, of type \texttt{root}, and empty content. Then, we run a pipeline composed from the following editors:

\begin{figure}
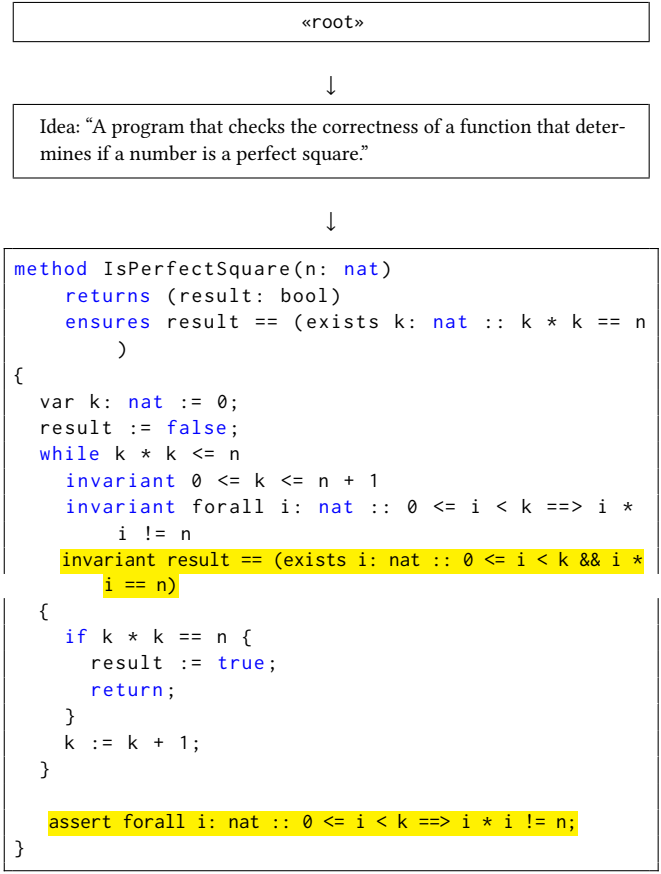

    \centering
    \footnotesize

\begin{mdframed}
\centering
\texttt{<<root>>}
\end{mdframed}

$$\downarrow$$

\begin{mdframed}
Idea: ``A program that checks the correctness of a function that determines
if a number is a perfect square.''
\end{mdframed}

$$\downarrow$$

\begin{Dafny}
method IsPerfectSquare(n: nat) 
    returns (result: bool)
    ensures result == (exists k: nat :: k * k == n)
{
  var k: nat := 0;
  result := false;
  while k * k <= n
    invariant 0 <= k <= n + 1
    invariant forall i: nat :: 0 <= i < k ==> i * i != n
   (*@ \hl{invariant result == (exists i: nat :: 0 <= i < k \&\& i * i == n)} @*)
  {
    if k * k == n {
      result := true;
      return;
    }
    k := k + 1;
  }

  (*@ \hl{assert forall i: nat :: 0 <= i < k ==> i * i != n;} @*)
}
\end{Dafny}
    
    \caption{Example taken from DafnySynth, of a sequence of edits that produces a fully verified method for training \tool{}. The root node at the top has no content, and is used to initialize the edit graph. The ``Idea Proposer'' creates nodes from the root node by asking an LLM to write a high-level idea of a verified program one might implement. Then, the ``Idea Implementer'' takes \texttt{idea} nodes as inputs, and again asks an LLM to write a small Dafny program related to that description. Here, it produces all lines except for the highlighted annotations (10 and 19). It checks that the resulting program is valid and that all of its annotations are accepted by Dafny, though initially the program does not fully verify. Then, an ``Annotator'' editor can take \texttt{program} nodes that don't fully verify, and attempt to add annotations to them. Here, GPT-4o adds an invariant and an assertion (\hl{highlighted} lines), resulting in a fully verified program. This new node can still be used further by other editors as the basis or larger, more complex programs, that can now call this method or extend it.}
    \label{fig:editors-example}
\end{figure}

\begin{description}
    \item[Idea Proposer: ] Takes the initial \texttt{root} node and existing nodes of type \texttt{idea}, and uses an LLM to sample a set of N new ``ideas'' (high-level natural language descriptions of a simple verified program one could write in Dafny), while avoiding the existing ideas in the graph. Figure~\ref{fig:editors-example} shows one simple idea sampled from GPT-4o as a proposer: checking if an integer is a perfect square.
    \item[Idea Implementer: ]Samples a set of \texttt{idea} nodes from the graph, and uses an LLM to write a simple, complete Dafny program, based on that idea. This yields \texttt{program} nodes, where the content is their source code. Given the idea above, GPT-4o implements the program shown in Figure~\ref{fig:editors-example}, at first without the highlited lines in yellow. This program is valid, although Dafny cannot yet prove the ensures clause. Since it can prove all of the existing invariants, we thus keep this program in the graph, saving the feedback from the verifier.
    \item[Annotator: ] Samples \texttt{program} nodes that don't yet verify, and attempts to add annotations to them, using an LLM. In the program in Figure~\ref{fig:editors-example}, given the existing program and an instruction to add annotations, GPT-4o adds the highlighted lines in yellow. In this case, Dafny can verify the new program (removing either of the annotations makes it fail). In general, we keep any new annotations as long as they can be proved by Dafny, regardless of whether the new program is already fully verified.
    \item[Change Proposer: ] Proposes and implements small patches to existing \texttt{program} nodes. We ask the LLM to either rewrite parts of the current program and/or add new code to it, aiming to produce a small patch that adds or modifies functionality. This allows the program to grow in complexity across edits, building on previous abstractions.
\end{description}

We thus run these annotators in sequence over the initial graph, allowing it to grow over time. This architecture is flexible, and new editors can be easily integrated. For instance, proposals for new ideas might come from translating existing programs from other languages, and edits can also attempt to mimick changes that human developers proposed in open-source repositories.

While this process arguably might not create fundamentally novel ideas for programs, the generated programs can still be useful given that they will be new programs \emph{in Dafny}. Our only concern is whether these programs will serve as useful training data for a tool like \tool{}. More broadly, we believe that a combination of symbolic verifiers, like Dafny's, and imperfect but very flexible generative models of code, gives a path forward to training AI assistants even prior to widespread adoption of a tool by developers.

In an initial exploration of this idea, we used \$10 in OpenAI API credits, querying the \texttt{gpt-4o-2024-08-06} model in all of the editors we described above. This gives us a dataset of 699 compilable Dafny programs --- over half of the raw size of DafnyBench (though many are derived from one another, so they are not unique). Of these, 46 are fully verified (Dafny proves \emph{all} specifications in all methods), whereas the remainder are partially verified (all of the annotations are accepted by Dafny, though Dafny cannot prove all specifications). We take all annotations as training examples, de-duplicating them to avoid copies of the same annotation that are simply carried over edits of a given program. Overall, DafnySynth contains 1107 annotations (678 invariants, 281 assertions, and 148 decreases clauses), of which 923 are syntactically unique --- each such annotation thus becomes one training example, once we pair it with a program where that annotation was removed.

\section{Experiments}
\label{sec:experiments}

\begin{table}
    \centering

\begin{tabular}{lcc}
\toprule
\textbf{Base model} & \textbf{Fine-tuning data} & \textbf{Success Rate} \\
\midrule
Llama 3.1 8B & - & 15.7\%\\
 & DB (25\%) & 15.7\%\\
 & DB (50\%) & 20.5\%\\
 & DB (100\%) & 15.7\%\\
 & DafnySynth & 27.7\%\\
 & DB (25\%) + DafnySynth & 41.0\%\\
 & DB (50\%) + DafnySynth & 43.4\%\\
 & DB (100\%) + DafnySynth & \textbf{50.6}\%\\
\midrule
CodeLlama 7B & - & 6.0\%\\
 & DB (25\%) & 20.5\%\\
 & DB (50\%) & 28.9\%\\
 & DB (100\%) & \textbf{39.8}\%\\
 & DafnySynth & 20.5\%\\
 & DB (25\%) + DafnySynth & 31.3\%\\
 & DB (50\%) + DafnySynth & 34.9\%\\
 & DB (100\%) + DafnySynth & \textbf{39.8}\%\\
\bottomrule
\end{tabular}

    \caption{Success rate of \tool{} in verifying a held-out set of 83 programs from DafnyBench, after annotations have been stripped. We use LLaMa 3.1 8B and CodeLlama 7B as base models, fine-tuned on various combinations of subsamples of DafnyBench and DafnySynth, our fully synthetic dataset. The best performing model is LLaMA 3.1 8B when trained on a combination of DafnyBench and DafnySynth.}
    \label{tab:result}
\end{table}

We implement \tool{} using the PyTorch \cite{pytorch} and vLLM libraries \cite{vllm}, and use it in combination with two recent open-weights LLMs hosted on Hugging Face \cite{hf}: LLaMa 3.1 8B \cite{llama} and CodeLlama 7B \cite{codellama}. We experiment both with the base models, as well as versions fine-tuned on various datasets we obtain from combinations of DafnySynth and DafnyBench.

To evaluate \tool{}, we split DafnyBench into a training set of 1000 Dafny methods, holding out the remaining 326 for test. We then strip annotations from the test methods, and measure the success rate of \tool{} at recovering annotations until the Dafny verifier accepts the program. Out of the 326 test methods, only 83 failed to verify after we removed the original annotations. We thus exclude the remaining programs from the evaluation, and focus on the 83 where Dafny needs hints to prove the specification.

For fine-tuning \tool{}, we use Low-Rank Adaptation (LoRA \cite{lora}), a lightweight fine-tuning method that requires significantly less GPU memory than full fine-tuning. All experiments were ran on a machine with one NVIDIA A100 80GB GPU, and 30 cores, used to parallelize calls to Dafny during greedy search. We use $r = \alpha = 128$, and train all models (regardless of their training set size) for $1500$ gradient steps (thus, all fine-tuning runs use the same amount of compute: we only vary their training dataset).

To construct training sets, we take subsamples of 0\%, 25\%, 50\% and 100\% of the 1000 training programs from DafnyBench (DB), and also vary whether we augment each of these DafnyBench subsamples with DafnySynth. This gives 8 variations of the training set (one of which is the ``trivial'' empty training set, corresponding to no fine-tuning). We then evaluate the success rate of \tool{}, when guided by a model fine-tuned in each of these datasets, with up to 5 iterations of the greedy search procedure from Section~\ref{sec:method}.

Table~\ref{tab:result} shows the results with both LLaMA 3.1 8B and CodeLlama 7B. We make the following observations:

\paragraph{Even small-scale fine-tuning can help \tool{} substantially}
Even with the order of a few thousand training examples in our datasets --- much less than typical for LLM training, an LLM-guided tool like \tool{} can enjoy substantial performance gains. The base LLaMA 3.1 8B model manages to verify 15.7\% of the programs in the test set --- after fine-tuning on DafnySynth alone, this number improves to 27.7\%. Similarly, for CodeLlama 7B, the base performance improves from only 6\% to 20.5\% after fine-tuning on DafnySynth. Moreover, we generally see gains as we increase the percentage of DafnyBench used for training (a trend that only breaks when \emph{not} using DafnySynth in combination with it when training LLaMa 3.1 8B).

\paragraph{DafnySynth provides high-quality training data}
Our best result is obtained in a combination of DafnySynth and DafnyBench, and using LLaMA 3.1 8B: using all of the available training data in both datasets yields a model with a success rate of 50.6\%. Interestingly, for LLaMA 3.1 8B, using DafnySynth alone is better than using DafnyBench for training, while for CodeLlama, DafnySynth leads to the same improvement as training on 250 programs from DafnyBench.

\paragraph{General code pre-training might not always help languages with smaller representation}
We do not see gains from using CodeLlama compared to the base Llama 3.1 8B model, despite it having been trained with a focus on code. Mainstream languages like Python will have a much larger representation in CodeLlama's training set, which possibly accounts for this fact. Pre-training on those languages does not seem to necessarily help with verifying Dafny programs, since the main feature we're interested in --- verification annotations --- is not present in popular languages. We note that this ability is still likely helpful in generating DafnySynth: while GPT-4o has not been trained on a substantial amount of Dafny code, having been exposed to millions of human-written programs in other languages seems to make it fairly proficient in \emph{implementing} Dafny code, even when it is not capable of helping Dafny verify it.

The performance of tools like \tool{} is highly driven by their training data: substantial gains can arise from fine-tuning on high-quality, task-relevant examples. In turn, the lack of available data can be potentially overcome via synthetic data generation, as we did in DafnySynth. Our results suggest that our pipeline, if refined and ran at scale, might be able to generate a large-scale, high-quality dataset for training LLMs to assist users in various tasks related to Dafny.

\section{Related Work}

Our work is most closely related to recent efforts using LLMs in program verification, in Dafny and other languages. Recent work in languages such as Isabelle (e.g. Baldur \cite{first2023baldur}), Coq (e.g. PALM \cite{lu2024proof}), and Lean (e.g. miniCodeProps \cite{lohn2024minicodeprops}) focus on \emph{proof synthesis} using LLMs, since proofs must be explicitly constructed in these languages. As is typically the case in deep learning, fine-tuning on domain-specific data dramatically boosts performance of these systems \cite{first2023baldur}. Coq and Isabelle have had a history of larger developments that allows data-driven approaches to benefit from already existing training data. For example, Isabelle's Archive of Formal proofs has more than 100k proofs contributed by users. While still relatively small compared to the most popular programming languages, these examples can already support a range of data-driven approaches for proving properties in these languages.

We build on DafnyBench \cite{loughridge2024dafnybench}, a recent dataset of Dafny programs collected from Github. DafnyBench is the largest Dafny dataset so far, and yet it has only 1326 programs. While we show that even small-scale fine-tuning on DafnyBench can already boost LLM performance, we believe that synthetic data generation, as has been applied in other domains \cite{zelikman2022star,poesia2024learning,lin2024lean}, can be a promising path for generating high-quality training data for Dafny.

\section{Conclusion}

We introduced \tool{}, an LLM-guided tool for automatically annotating Dafny methods.
While our initial focus was in generating annotations, we note that these alone are often not the main difficulty of writing verified programs in Dafny. \tool{} assumes that programs come with formal specifications, while a challenge for users is often in formally describing the desired behavior. We believe that LLMs have a great potential to help in that step, as well, due to their understanding of natural language and of coding patterns that are common in human-written code and thus might help disambiguate vague specifications given in natural language.

Also, many of the DafnyBench programs already come with helpful lemmas, which \tool{} is often able to \emph{use} in the annotations it produces. In actuality, these lemmas must also be introduced by users during development, which shows another opportunity for assistance that we plan to explore in future work.

LLMs combined with symbolic methods have an enormous potential to help users write verified programs. The lack of high-quality training data at scale for these tasks can be a challenge for leveraging the full potential of LLMs, which tend to require sizeable training datasets to perform well. However, synthetic generation methods, such as the DafnySynth pipeline, can be used to generate relevant training data to help with all activities involved in verified programming: from generating specifications, proposing lemmas, producing implementations, as well as working to aid the verifier with annotations. These tools might drastically lower the cost of producing verified software in the future --- software produced either interactively by users, by translating from unverified languages, or even by synthesizing entire implementations. In turn, as proof automation improves, we expect formal verification to become attractive for a wider audience, generally improving the reliability of the software we depend on.

\section*{Acknowledgements}
We thank Dennis Du, Anastasiya Kravchuk-Kirilyuk and Simon Henniger for discussions during meetings. Chloe Loughridge was partially supported by the Harvard College Research Program (HCRP) through the Harvard College Office of Undergraduate Research and Fellowships. Nada Amin was partially supported by NSF Award 2303983. Gabriel Poesia was supported by a Stanford Graduate Interdisciplinary Fellowship (SIGF).

\bibliography{acmart}

\begin{thebibliography}{10}

\bibitem{coq}
Bruno Barras, Samuel Boutin, Cristina Cornes, Judica{\"e}l Courant, Yann
  Coscoy, David Delahaye, Daniel de~Rauglaudre, Jean-Christophe Filli{\^a}tre,
  Eduardo Gim{\'e}nez, Hugo Herbelin, et~al.
\newblock The coq proof assistant reference manual.
\newblock {\em INRIA, version}, 6(11), 1999.

\bibitem{llama}
Abhimanyu Dubey, Abhinav Jauhri, Abhinav Pandey, Abhishek Kadian, Ahmad
  Al-Dahle, Aiesha Letman, Akhil Mathur, Alan Schelten, Amy Yang, Angela Fan,
  et~al.
\newblock The llama 3 herd of models.
\newblock {\em arXiv preprint arXiv:2407.21783}, 2024.

\bibitem{first2023baldur}
Emily First, Markus~N Rabe, Talia Ringer, and Yuriy Brun.
\newblock Baldur: Whole-proof generation and repair with large language models.
\newblock In {\em Proceedings of the 31st ACM Joint European Software
  Engineering Conference and Symposium on the Foundations of Software
  Engineering}, pages 1229--1241, 2023.

\bibitem{lora}
Edward~J Hu, Yelong Shen, Phillip Wallis, Zeyuan Allen-Zhu, Yuanzhi Li, Shean
  Wang, Lu~Wang, and Weizhu Chen.
\newblock Lora: Low-rank adaptation of large language models.
\newblock {\em arXiv preprint arXiv:2106.09685}, 2021.

\bibitem{krasner2021cost}
Herb Krasner.
\newblock The cost of poor software quality in the us: A 2020 report.
\newblock {\em Proc. Consortium Inf. Softw. QualityTM (CISQTM)}, 2, 2021.

\bibitem{vllm}
Woosuk Kwon, Zhuohan Li, Siyuan Zhuang, Ying Sheng, Lianmin Zheng, Cody~Hao Yu,
  Joseph Gonzalez, Hao Zhang, and Ion Stoica.
\newblock Efficient memory management for large language model serving with
  pagedattention.
\newblock In {\em Proceedings of the 29th Symposium on Operating Systems
  Principles}, pages 611--626, 2023.

\bibitem{dafny}
K.~Rustan~M. Leino.
\newblock Dafny: an automatic program verifier for functional correctness.
\newblock In {\em Proceedings of the 16th International Conference on Logic for
  Programming, Artificial Intelligence, and Reasoning}, LPAR'10, page
  348–370, Berlin, Heidelberg, 2010. Springer-Verlag.

\bibitem{lin2024lean}
Haohan Lin, Zhiqing Sun, Yiming Yang, and Sean Welleck.
\newblock Lean-star: Learning to interleave thinking and proving.
\newblock {\em arXiv preprint arXiv:2407.10040}, 2024.

\bibitem{lohn2024minicodeprops}
Evan Lohn and Sean Welleck.
\newblock minicodeprops: a minimal benchmark for proving code properties.
\newblock {\em arXiv preprint arXiv:2406.11915}, 2024.

\bibitem{loughridge2024dafnybench}
Chloe Loughridge, Qinyi Sun, Seth Ahrenbach, Federico Cassano, Chuyue Sun, Ying
  Sheng, Anish Mudide, Md~Rakib~Hossain Misu, Nada Amin, and Max Tegmark.
\newblock Dafnybench: A benchmark for formal software verification.
\newblock {\em arXiv preprint arXiv:2406.08467}, 2024.

\bibitem{lu2024proof}
Minghai Lu, Benjamin Delaware, and Tianyi Zhang.
\newblock Proof automation with large language models.
\newblock {\em arXiv preprint arXiv:2409.14274}, 2024.

\bibitem{isabelle}
Lawrence~C Paulson.
\newblock {\em Isabelle: A generic theorem prover}.
\newblock Springer, 1994.

\bibitem{poesia2024learning}
Gabriel Poesia, David Broman, Nick Haber, and Noah~D Goodman.
\newblock Learning formal mathematics from intrinsic motivation.
\newblock {\em arXiv preprint arXiv:2407.00695}, 2024.

\bibitem{codellama}
Baptiste Roziere, Jonas Gehring, Fabian Gloeckle, Sten Sootla, Itai Gat,
  Xiaoqing~Ellen Tan, Yossi Adi, Jingyu Liu, Romain Sauvestre, Tal Remez,
  et~al.
\newblock Code llama: Open foundation models for code.
\newblock {\em arXiv preprint arXiv:2308.12950}, 2023.

\bibitem{hf}
Thomas Wolf, Lysandre Debut, Victor Sanh, Julien Chaumond, Clement Delangue,
  Anthony Moi, Pierric Cistac, Tim Rault, R{\'e}mi Louf, Morgan Funtowicz,
  et~al.
\newblock Transformers: State-of-the-art natural language processing.
\newblock In {\em Proceedings of the 2020 conference on empirical methods in
  natural language processing: system demonstrations}, pages 38--45, 2020.

\bibitem{pytorch}
Peng Wu.
\newblock Pytorch 2.0: The journey to bringing compiler technologies to the
  core of pytorch (keynote).
\newblock In {\em Proceedings of the 21st ACM/IEEE International Symposium on
  Code Generation and Optimization}, pages 1--1, 2023.

\bibitem{zelikman2022star}
Eric Zelikman, Yuhuai Wu, Jesse Mu, and Noah Goodman.
\newblock Star: Bootstrapping reasoning with reasoning.
\newblock {\em Advances in Neural Information Processing Systems},
  35:15476--15488, 2022.

\end{thebibliography}
\bibliographystyle{plain}

\end{document}